# Dynamic Regressor Extension and Mixing-based Redesign of Adaptive Observer for Affine Systems

Mehdi Tavan

*Abstract*— **The dynamic regressor extension and mixing procedure is employed to redesign a conventional adaptive observer algorithm for affine systems. A reduced-order observer is designed without the construction of the state transition matrix. The dynamics of the regressor are redesigned to incorporate feedback from its extension, transforming the regressor dynamics into a perturbed damped nonlinear oscillator form. This introduces some flexibility in reducing the degradation of parameter convergence due to the lack of the transition matrix and in enhancing the excitation property of the extension matrix.**

***Index Terms***—**Adaptive observers, affine systems, dynamic regressors.**

## I. INTRODUCTION

In 1977, Kreisselmeier proposed a regressor extension in [1], which generates a positive definite regressor from a regressor with the persistency of excitation (PE) property [2]. The desire to achieve high convergence rate in parameter estimation has once again attracted researchers' attention to employing this method of regressor expansion. In [3], Marino and Tomei's adaptive observer [4] is equipped with the Kreisselmeier regressor extension to enhance parameter convergence in scenarios where the PE property is lacking. The mixing process proposed in [5] has been applied to the Kreisselmeier regressor extension to form generalized parameter estimation-based (GPEB) observers for affine systems in state [6] and in both state and parameter [7]. The resulting dynamic regressor extension and mixing (DREM) process relaxes the parametric convergence condition to a non-square integrability condition. The *key step* in the design of GPEB observers is constructing an asymptotically stable fundamental matrix that acts as the state transition matrix. As a result, the degree of freedom of the observer increases. However, an additional issue emerges when the system's state or parameters change after the fundamental matrix approaches zero. In such cases, a reset mechanism is required to revert it to the identity matrix $I$, as done in [7].

In this note, a DREM-based redesign of the adaptive observer in [8] is proposed via Kreisselmeier regressor extension to relax the parametric convergence to a non-square integrability condition. A reduced-order observer is obtained without the need to construct a fundamental matrix. A concern about the proposed algorithm is that the parameter estimation process will be disturbed by a perturbation dependent on the state estimation error, as seen in the algorithm proposed in [8] (refer to Equations (14) and (15) in [8]). To reduce this effect, the dynamics of the regressor are modified to leverage *the joy of feedback* from its extension, forming the dynamic regressor and its extension into a perturbed damped nonlinear oscillator. This can provide two primary advantages without increasing the observer's dimensionality. Firstly, the state estimation error's impact on parameter estimation dynamics can be arbitrarily minimized. Secondly, the regressor dynamics can exhibit an oscillatory transient response. Both features are crucial when the parameter regressor experiences weak excitation properties.

*Notation*: For $\mathcal{J} \in \mathbb{R}^{\mathrm{p}\times\mathrm{p}}$, $\mathrm{adj}[\mathcal{J}]$ and $\det[\mathcal{J}]$ represent the adjunct matrix and the determinant of $\mathcal{J}$, respectively. $I \in \mathbb{R}^{\mathrm{p}\times\mathrm{p}}$ is the identity matrix. For general mappings $V: \mathbb{R}^{\mathrm{x}} \times \mathbb{R}_{\geq 0} \to \mathbb{R}$, we define $\nabla_x V(x,t) := \partial V(x,t)/\partial x$.

## II. PROBLEM STATEMENT

In this study we are interested in observer design for a class of nonlinear systems whose dynamics can be represented in suitable co-ordinates by equations of the following form

$$\dot{x} = \mathcal{A}(u, y)x + \Omega(u, y)\theta + \mathscr{b}(u, y), \tag{1}$$
$$y = \mathcal{C}(u)x + \Psi(u)\theta, \tag{2}$$

with $x \in \mathbb{R}^{\mathrm{x}}$ is the state vector, $u \in \mathbb{R}^{\mathrm{u}}$ is a measured exogenous input, $y \in \mathbb{R}^{\mathrm{y}}$ is the output vector, $\theta \in \mathbb{R}^{\mathrm{p}}$ is an unknown constant vector, $\mathscr{b}: \mathbb{R}^{\mathrm{u}} \times \mathbb{R}^{\mathrm{y}} \to \mathbb{R}^{\mathrm{x}}$, the mappings $\mathcal{A}: \mathbb{R}^{\mathrm{u}} \times \mathbb{R}^{\mathrm{y}} \to \mathbb{R}^{\mathrm{x}\times\mathrm{x}}$ and $\mathcal{C}: \mathbb{R}^{\mathrm{u}} \to \mathbb{R}^{\mathrm{y}\times\mathrm{x}}$ are continuous and bounded, and the mappings $\Omega: \mathbb{R}^{\mathrm{u}} \times \mathbb{R}^{\mathrm{y}} \to \mathbb{R}^{\mathrm{x}\times\mathrm{p}}$ and $\Psi: \mathbb{R}^{\mathrm{u}} \to \mathbb{R}^{\mathrm{y}\times\mathrm{p}}$ are bounded. The properties of the foregoing mappings hold uniformly for any $u$ and $y$.

Following [4] and [8], to facilitate transforming the system dynamics into a proper adaptive observer form, we consider the following filtered transformation

$$z(t) := x(t) - \Upsilon(t)\theta, \tag{3}$$

where $\Upsilon: \mathbb{R}_{\geq 0} \to \mathbb{R}^{\mathrm{x}\times\mathrm{p}}$ is an auxiliary filter, whose dynamics are to be specified. Consequently, the system (1)-(2) can be represented in $z$-co-ordinates as

$$\begin{aligned}\dot{z} &= \mathcal{A}(u, y)z + \mathscr{b}(u, y)\\ &\quad+\big(\mathcal{A}(u, y)\Upsilon(t) + \Omega(u, y) - \dot{\Upsilon}\big)\theta,\end{aligned} \tag{4}$$
$$y = \mathcal{C}(u)z + \Xi(u, \Upsilon)\theta, \tag{5}$$

with

$$\Xi(u, \Upsilon) := \mathcal{C}(u)\Upsilon(t) + \Psi(u). \tag{6}$$

Mehdi Tavan is with the Department of Electrical Engineering, Nour Branch, Islamic Azad University, Nour, Iran (e-mail: m.tavan@srbiau.ac.ir).

The problem is to design an adaptive observer to form the estimated state $\hat{z}(t)$ and the estimated parameter $\hat{\theta}(t)$ such that the estimation errors

$$\bar{z}(t) := z(t) - \hat{z}(t), \tag{7}$$
$$\bar{\theta}(t) := \theta - \hat{\theta}(t), \tag{8}$$

converge to zero, allowing for an asymptotic estimate of the state $x(t)$ to be admissible by

$$\hat{x}(t) = \hat{z}(t) + \Upsilon(t)\hat{\theta}(t), \tag{9}$$

for the bounded trajectories of the filter $\Upsilon(t)$.

## III. Design procedure

We make the following assumptions to design a DREM-based adaptive observer for the system addressed in Section II.

*Assumption 1* There exists a globally Lipschitz mapping $\Gamma: \mathbb{R}^{\mathfrak{u}} \times \mathbb{R}^{\mathfrak{y}} \to \mathbb{R}^{\mathfrak{x}\times\mathfrak{y}}$ such that the mapping

$$\Lambda(u, y) := \mathcal{A}(u, y) - \Gamma(u, y)\mathcal{C}(u), \tag{10}$$

is continuous and bounded, and the trajectories of the dynamics

$$\dot{\Upsilon} = \Lambda(u, y)\Upsilon + \Omega(u, y) - \Gamma(u, y)\Psi(u), \tag{11}$$

remains bounded, uniformly in $u$, $y$, and $\Upsilon(0) \in \mathbb{R}^{\mathfrak{x}\times\mathfrak{y}}$.

*Assumption 2* There exists a Lyapunov function $V: \mathbb{R}^{\mathfrak{x}} \times \mathbb{R}_{\geq 0} \to \mathbb{R}_{\geq 0}$ whose time derivative along the trajectories

$$\dot{\bar{z}} = \Lambda(u, y)\bar{z}, \tag{12}$$

satisfies

$$\dot{V} \leq -|\mathcal{C}(u)\bar{z}|^2. \tag{13}$$

*Assumption 3* The origin of the dynamical system (12) is globally exponentially stable (GES), uniformly in $u$, $y$, and $\bar{z}(0) \in \mathbb{R}^{\mathfrak{x}}$.

Although Assumption 1 is stated in general terms, the mapping $\Gamma(u, y)$ that guaranties Assumption 3 for the dynamical system (12) also satisfies Assumption 1 by invoking the converse Lyapunov theorem (such as Theorem 4.14 in [9], Lemma 1 in [10], and Theorem 3.3 in [11]), and the arguments known as total stability (such as Lemma 9.2, 9.3, Exercise 9.21, the last paragraph on page 349 in [9], and Theorem 24.1 in [12]). In fact, Assumption 1 aims to leave room for systems in the form of (12) with weaker stability properties that still result in bounded trajectories for the dynamical system (11). The proposition 3 and Remark 4 in the next section discuss the feasibility of the assumption.

*Proposition 1* Consider the system (1)-(2) that confirms Assumption 1, in conjunction with (11) and the observer

$$\dot{\hat{z}} = \Lambda(u, y)\hat{z} + \Gamma(u, y)y + b(u, y), \tag{14}$$
$$\dot{\hat{\theta}} = \lambda\, \mathrm{adj}[\Phi(t)]\left(\mathcal{Y}(t) - \Phi(t)\hat{\theta}\right), \tag{15}$$
$$\dot{\mathcal{Y}} = -\kappa\mathcal{Y} + \Xi^{\top}(u, \Upsilon)(y - \mathcal{C}(u)\hat{z}), \tag{16}$$
$$\dot{\Phi} = -\kappa\Phi + \Xi^{\top}(u, \Upsilon)\Xi(u, \Upsilon), \Phi(0) = \Phi^{\top}(0) \geq 0 \tag{17}$$

where $\lambda$ and $\kappa$ are arbitrary positive constants. Suppose that Assumption 2 holds and

$$\delta(t) := \sqrt{\det[\Phi(t)]} \notin \mathcal{L}_2. \tag{18}$$

Then $\bar{z}$ and $\bar{\theta}$ are bounded. Moreover, $\bar{\theta}$ asymptotically converges to zero if $\delta(t) \in PE$, that is,

$$\int_t^{t+T} \delta^2(\tau)\, d\tau \geq \rho, \tag{19}$$

for some $T, \rho > 0$ and for all $t \in \mathbb{R}_{\geq 0}$. Suppose that Assumption 3 holds. Then $\bar{z}$ converges to zero exponentially, $\bar{\theta}$ converges to zero asymptotically under condition (18) and exponentially under condition (19).

*Proof 1* To begin with, note that multiplying (5) from the left with $\Xi^{\top}(u, \Upsilon)$ yields

$$\Xi^{\top}(u, \Upsilon)(y - \mathcal{C}(u)z) = \Xi^{\top}(u, \Upsilon)\Xi(u, \Upsilon)\theta. \tag{20}$$

This implies, by linearity, the matrixes $\mathcal{Y}(t)$ and $\Phi(t)$, generated respectively by (16) and (17), are related by

$$\mathcal{Y}(t) - \epsilon(t) = \Phi(t)\theta, \tag{21}$$

where $\epsilon(t)$ is generated by

$$\dot{\epsilon} = -\kappa\epsilon + \Xi^{\top}(u, \Upsilon)\mathcal{C}(u)\bar{z}. \tag{22}$$

Since $\Phi(t) = \Phi^{\top}(t) \geq 0$ for all time [3], it can be concluded that $\delta(t)$ is well defined for all time. Now, multiplying (21) from the left with $\lambda\, \mathrm{adj}[\Phi]$ yields

$$\lambda\, \mathrm{adj}[\Phi(t)]\left(\mathcal{Y}(t) - \epsilon(t)\right) = \lambda\delta^2(t)\theta, \tag{23}$$

where the facts $\det(\mathcal{J})\, I = \mathrm{adj}(\mathcal{J})\, \mathcal{J}$ for any matrix $\mathcal{J} \in \mathbb{R}^{\mathfrak{p}\times\mathfrak{p}}$ has been used to derive the above equality. Differentiating (8), replacing (15) therein, and using (23), we get

$$\dot{\bar{\theta}} = -\lambda\delta^2(t)\bar{\theta} + \varepsilon(t), \tag{24}$$

with

$$\varepsilon(t) := -\lambda\, \mathrm{adj}[\Phi(t)]\, \epsilon(t). \tag{25}$$

By invoking Assumption 1, we can conclude that the mapping $\Xi(u, \Upsilon)$, given by (6), is bounded, and this property holds uniformly in $u$ and $\Upsilon$. Consequently, this implies that $\Phi(t)$, given by (17), is also bounded. Next, differentiating (7) and using (4), (10) and (14) implies that (11) fixes the dynamics of $\bar{z}(t)$ to (12). Furthermore, either Assumption 2 or Assumption 3 ensures that $\bar{z}(t)$ is bounded, which in turn implies the boundedness of the signals $\mathcal{Y}$, $\epsilon$, and $\varepsilon$, given by (16), (22), and (25), respectively. We can also conclude that $\dot{\Phi}$ and $\dot{\epsilon}$ belong to $\mathcal{L}_\infty$, which leads to $\dot{\varepsilon} \in \mathcal{L}_\infty$. Now, notice that from Assumption 2 it follows that $\mathcal{C}(u)\bar{z} \in \mathcal{L}_2$, which consequently leads to $\epsilon \in \mathcal{L}_2$ and $\varepsilon \in \mathcal{L}_2$. As a result, the boundedness of the estimation error $\bar{\theta}$, generated by (24), is guaranteed by Proposition 1 in [13] under condition (18). On the other hand, by invoking standard results on cascaded systems from [14], we can conclude that the origin $(\bar{z}, \epsilon) = 0$ is a uniformly globally exponentially stable equilibrium for the cascaded system described by (12) and (22) under Assumption 3. This means that $\varepsilon(t)$ tends to zero exponentially, implying that $\varepsilon \in \mathcal{L}_1$. Therefore, based on Lemma 1 in [15], condition (18) ensures that the origin of the perturbed system (24) is uniformly globally asymptotically stable. By setting $\varepsilon(t) = 0$ in (24), the resulting unperturbed system is uniformly globally exponentially stable under condition (19) as shown in [16]. As a result, the converse Lyapunov theorem (such as Theorem 4.14 in [9], Lemma 1 in [10], and Theorem 3.3 in [11]) guarantees the existence of a *proper* Lyapunov function for the unperturbed system that satisfies the conditions of Proposition 2.3 in [14]. Consequently, the cascade connection between (24) and the dynamics of $(\epsilon, \bar{z})$ preserves the uniformly globally exponential stability property at the origin under condition (19) and Assumption 3. Since $\dot{\varepsilon} \in \mathcal{L}_\infty$, the property $\varepsilon \in \mathcal{L}_2$, resulting from Assumption 2, implies that $\varepsilon(t)$ asymptotically converges to zero by invoking the generalized Barbalat's lemma [17]. Therefore, under condition (19), the equilibrium $\bar{\theta} = 0$ is uniformly globally asymptotically stable by Lemma 9.4 and invoking Lemma 9.6 both in [9]. This completes the proof. □

*Remark 1* The proposed design procedure simplifies the degree of freedom compared to existing GPEB observer designs, such as Proposition 1 in [7], by eliminating the need to construct an asymptotic stable fundamental matrix acts as the state transition matrix. The main concern regarding the proposed algorithm in Proposition 1 is the perturbation term $\varepsilon(t)$, given by equation (25), which increases with the rise in $\lambda$ to enhance the transient response of the parameter estimation error dynamics (24). The same concern exists for the observer proposed in [8]. One of the functions of the fundamental matrix in the GPEB observer is to circumvent this concern. However, the fundamental matrix tending toward zero poses an obstacle to completing this task. Hence, a mechanism is needed to alert changes in the system's state or parameters and, in reaction, reset the fundamental matrix to the identity matrix $I$, as done in [7].

*Remark 2* It is well known from [18] that the mixing procedure reflected in the adaptation law (15) offers several advantages over the traditional gradient method, including a relaxed convergence condition as outlined in (18). However, a concern arises when the regressor extension scheme (17) is actuated by a dynamic regressor $\Xi$ that lacks persistency of excitation. In this case, (17) provides a positive semi-definite matrix, and the updating process in (15) will face a disruption because the determinant of $\Phi$ is zero. From a robustness viewpoint, it may be better to replace the term $\mathrm{adj}[\Phi]$ with $\delta\,\mathrm{adj}[\Phi]$ in (15) to completely stop the updating process when $\det[\Phi]$ becomes zero. Notice that replacing the term $\mathrm{adj}[\Phi(t)]$ with the identity matrix $I$ may provide parameter convergence commensurate with the rank of $\Phi$.

## IV. Redesign Procedure

Considering (22), it is clear that the transient response of $\epsilon$, and consequently $\varepsilon$, is a function of the initial condition of $\bar{z}$. To mitigate the side effects of the initial condition of $\bar{z}$ on the performance of the parameter estimation error, the dynamics of the filter $\Upsilon$ and the estimator $\hat{z}$ are redesigned using the following assumptions.

*Assumption 4* The gradient of the known Lyapunov function $V(\bar{z}, t)$, specified in Assumption 2, along the trajectories (12) satisfies

$$\nabla_{\bar{z}} V = \bar{z}^\top \mathrm{P}(t), \tag{26}$$

for some continuous mapping $\mathrm{P} = \mathrm{P}^\top : \mathbb{R}_{\geq 0} \to \mathbb{R}^{\mathfrak{r}\times\mathfrak{r}}$, which is bounded and invertible uniformly in $t$.

*Assumption 5* The bondedness of $\Upsilon$-trajectories, assumed in Assumption 1, is preserved when its dynamics incorporate feedback from (17) as follow

$$\begin{aligned}\dot{\Upsilon} = \Lambda(u, y)\Upsilon + \Omega(u, y) - \Gamma(u, y)\Psi(u) - \\ \rho\mathcal{T}(u, \Upsilon)\Phi,\end{aligned} \tag{27}$$

where $\rho > 0$ is an arbitrary feedback gain and

$$\mathcal{T}(u, \Upsilon) \coloneqq \mathrm{P}^{-1}(t)\mathcal{C}^\top(u)\Xi(u, \Upsilon). \tag{28}$$

*Proposition 2* Consider the system (1)-(2) that confirms Assumptions 4 and 5, in conjunction with (15), (16), (17), (27) and the observer

$$\begin{aligned}\dot{\hat{z}} = \Lambda(u, y)\hat{z} + \Gamma(u, y)y + b(u, y) + \\ \rho\mathcal{T}(u, \Upsilon)\mathcal{Y}(t).\end{aligned} \tag{29}$$

Suppose that Assumption 2 holds. Then $\bar{z}$ and $\bar{\theta}$ are bounded if condition (18) is satisfied. Moreover, $\bar{\theta}$ asymptotically converges to zero if condition (19) is also satisfied. Suppose that Assumption 3 holds. Then $\bar{z}$ converges to zero exponentially, $\bar{\theta}$ converges to zero asymptotically under condition (18) and exponentially under condition (19).

*Proof 2* To begin with, note that since (15), (16), and (17) are repeated in Proposition 2, the dynamics of $\bar{\theta}$ and $\epsilon$ remain as described in (24) and (22), respectively. By inserting (27) in (4), and using (29) and (21), the dynamics of $\bar{z}$ and $\epsilon$ take the following damped nonlinear form

$$\dot{\bar{z}} = \Lambda(u, y)\bar{z} - \rho\mathcal{T}(u, \Upsilon)\epsilon, \tag{30}$$

$$\dot{\epsilon} = \Xi^\top(u, \Upsilon)\mathcal{C}(u)\bar{z} - \kappa\epsilon. \tag{31}$$

To establish the stability property of the aforementioned dynamical system, we consider the following Lyapunov function

$$V_0(\bar{z}, \epsilon, t) \coloneqq V(\bar{z}, t) + \frac{\rho}{2}|\epsilon|^2, \tag{32}$$

where $V(.)$ is defined in Assumptions 2 and 4. The time derivative of the Lyapunov function (32) along the trajectories (30) and (31) is

$$\begin{aligned}\dot{V}_0 &= \dot{V} - \nabla_{\bar{z}} V(\rho\mathcal{T}\epsilon) + \rho\dot{\epsilon}^\top\epsilon \\ &\leq -|\mathcal{C}(u)\bar{z}|^2 - \kappa\rho|\epsilon|^2,\end{aligned} \tag{33}$$

where we have used (26) and (28) to obtain the above bound. As a result, we obtain $\bar{z} \in \mathcal{L}_\infty$, $\mathcal{C}(u)\bar{z} \in \mathcal{L}_2$ and $\epsilon \in \mathcal{L}_\infty \cap \mathcal{L}_2$. Also, from Assumption 5 we deduce that $\Upsilon$ and $\Phi$ are bounded. Consequently, $\dot{\bar{z}} \in \mathcal{L}_\infty$ and $\dot{\epsilon} \in \mathcal{L}_\infty$ can be obtained from (30) and (31). Now, $\bar{\theta} \in \mathcal{L}_\infty$ can be concluded from $\epsilon \in \mathcal{L}_2$ and $\delta(t) \notin \mathcal{L}_2$ by invoking Proposition 1 in [13]. Additionally, since $\dot{\epsilon} \in \mathcal{L}_\infty$ and $\epsilon \in \mathcal{L}_2$, the asymptotic convergence of $\bar{\theta}$ to zero can be concluded under condition (19), following the same arguments at the end of Proof 1. Now, note that the dynamical system (30)-(31) under Assumption 3 can be seen as a nominal system with a uniformly GES equilibrium point at $(\bar{z}, \epsilon) = 0$, which is perturbed by a bounded perturbation dependent on the square-integrable signals $\epsilon$ and $\mathcal{C}(u)\bar{z} \in$. As a result, the origin $(\bar{z}, \epsilon) = 0$ of the perturbed system (30)-(31) remains uniformly GES by invoking Lemma 4 in [10]. As a result, $\epsilon \in \mathcal{L}_1$ and the asymptotic convergence of $\bar{\theta}$ is guaranteed in the perturbed system (24) under condition (18) by Lemma 1 in [15]. Once again, note that the system (30)-(31) and (24) can be seen as the following nominal system

$$\dot{\bar{z}} = \Lambda(u, y)\bar{z}, \dot{\epsilon} = -\kappa\epsilon, \dot{\bar{\theta}} = -\lambda\delta^2(t)\bar{\theta}, \tag{34}$$

which is perturbed by bounded terms dependent on square-integrable signals $\epsilon$ and $\mathcal{C}(u)\bar{z}$. According to Assumption 3 and condition (19), the nominal system (34) has a uniformly GES equilibrium point at $(\bar{z}, \epsilon, \bar{\theta}) = 0$. Lemma 4 in [10] ensures that this stability property is retained for the systems described by (30)-(31) and (24). □

*Remark 3* Note that, for a well-defined term of $\mathcal{T}(u, \Upsilon)$, increasing $\rho$ in the damped nonlinear system (30)-(31) can reduce the overshoot of $\epsilon$ due to an initial condition $\bar{z}$ (see the example in [19], page 294). Consequently, the transient response of the parameter estimation error $\bar{\theta}$, generated by (24), can be enhanced for a given value of $\lambda$. This feature could be instrumental when the excitation property of the dynamic regressor diminishes over time.

Similar to Assumption 1, Assumption 5 is stated in general terms to leave room for the dynamical systems in the form of (12) with any type of stability that satisfies Assumptions 4 and 5. The following proposition demonstrates the feasibility of Assumptions 1 and 5.

*Proposition 3* Assume the mapping $\Gamma(u, y)$ is such that the dynamical system (12) satisfies Assumptions 2, 3 and 4. Then the dynamic regressor (27), in conjunction with (17), satisfies Assumption 5 for

$$\kappa > \rho\frac{\mathrm{p}}{4}\|\Psi(u)\|^2. \tag{35}$$

*Proof 3* To begin with, let's define the variables

$$v_i(t) := \Upsilon(t)\iota_i, \tag{36}$$

$$\varphi_i(t) := \Phi(t)\iota_i, \tag{37}$$

where $\iota_i \in \mathbb{R}^{\mathrm{p}}$ is a unit vector, whose $i$-th entry equals 1. Differentiating (36) and (37), and using (27) and (17), the dynamics of these variables can be written as the following perturbed system

$$\dot{v}_i = \Lambda(u, y)v_i - \rho\mathcal{T}(u, \Upsilon)\varphi_i + d(u, y), \tag{38}$$

$$\dot{\varphi}_i = \Xi^{\top}(u, \Upsilon)\mathcal{C}(u)v_i - \kappa\varphi_i + p(u, \Upsilon) + q(u), \tag{39}$$

where the perturbation terms are given by

$$d(u, y) := \Omega(u, y)\iota_i - \Gamma(u, y)\Psi(u)\iota_i, \tag{40}$$

$$p(u, \Upsilon) := \Upsilon^{\top}\mathcal{C}^{\top}(u)\Psi(u)\iota_i, \tag{41}$$

$$q(u) := \Psi^{\top}(u)\Psi(u)\iota_i. \tag{42}$$

Now, consider the Lyapunov function

$$V_0(\Phi, \Upsilon, t) := \sum_{i=1}^{\mathrm{p}}\left(V(v_i, t) + \frac{\rho}{2}|\varphi_i|^2\right), \tag{43}$$

whose time derivative along the trajectories of the unperturbed system, *i.e.*, the system (38)-(39) with $d(.) = p(.) = q(.) = 0$, is

$$\begin{aligned}\dot{V}_0 &= \textstyle\sum_{i=1}^{\mathrm{p}}\left(\dot{V} - \nabla_{v_i}V(\rho\mathcal{T}\varphi_i) + \rho\dot{\varphi}_i^{\top}\varphi_i\right) \\ &\leq -\textstyle\sum_{i=1}^{\mathrm{p}}(|\mathcal{C}v_i|^2 + \kappa\rho|\varphi_i|^2),\end{aligned} \tag{44}$$

where we have used (26) and (28) to obtain the above bound. As a result, we obtain $v_i \in \mathcal{L}_\infty$, $\mathcal{C}v_i \in \mathcal{L}_2$ and $\varphi_i \in \mathcal{L}_\infty \cap \mathcal{L}_2$ for any $\iota_i$-direction. As a result, $\Upsilon$ is bounded, and consequently $\Xi$ and $\mathcal{T}$ are bounded for the unperturbed system (38)-(39). By respectively substituting $v_i$ and $\varphi_i$ for $\bar{z}$ and $\epsilon$ in (30)-(31), and invoking the similar arguments to those in Proof 2, it can be concluded that the system (38)-(39) with zero perturbation terms has a uniformly GES equilibrium point at $(v_i, \varphi_i) = 0$. Now, notice that the time derivative of the Lyapunov function $V_0$ along the trajectories of the system (38)-(39) with $d(.) = q(.) = 0$ satisfies

$$\begin{aligned}\dot{V}_0 \leq -\textstyle\sum_{i=1}^{\mathrm{p}}(&|\mathcal{C}v_i|^2 + \kappa\rho|\varphi_i|^2 - \\ &\rho\|\Psi\||\mathcal{C}\Upsilon\iota||\varphi_i|),\end{aligned} \tag{45}$$

where

$$\iota := \frac{\varphi_i}{|\varphi_i|}, \tag{46}$$

and we have used

$$\iota_i^{\top}\Psi^{\top}(u)\mathcal{C}(u)\Upsilon\varphi_i \leq |\Psi(u)\iota_i||\mathcal{C}(u)\Upsilon\iota||\varphi_i|, \tag{47}$$

to get the bound (44). Using Young's inequality with the factor $2(1-\varrho)/\mathrm{p}$, where $\varrho$ is to be specified, and the fact that

$$\sum_{i=1}^{\mathrm{p}}\left(|\mathcal{C}(u)\Upsilon\iota_i|^2 - \frac{1}{\mathrm{p}}|\mathcal{C}(u)\Upsilon\iota|^2\right) \geq 0, \tag{48}$$

we obtain

$$\dot{V}_1 \leq -\sum_{i=1}^{\mathrm{p}}\left(\varrho|\mathcal{C}v_i|^2 + \rho\left(\kappa - \rho\frac{\mathrm{p}\|\Psi\|^2}{4(1-\varrho)}\right)|\varphi_i|^2\right). \tag{49}$$

Now, notice that for any given $\kappa$, which satisfied (35), there exists a $\varrho \in (0,1)$ for (49) such that, for any $\iota_i$-direction, $v_i \in \mathcal{L}_\infty$, $\mathcal{C}v_i \in \mathcal{L}_2$ and $\varphi_i \in \mathcal{L}_\infty \cap \mathcal{L}_2$. As a result, $\Upsilon$ is bounded, and consequently $\Xi$ and $\mathcal{T}$ are bounded. Additionally, since $\mathcal{C}v_i \in \mathcal{L}_2$ holds for any $\iota_i$-direction it can be concluded that $p(u, \Upsilon) \in \mathcal{L}_2$. As a result, the equilibrium point $(v_i, \varphi_i) = 0$ of the perturbed system (38)-(39) with $d(.) = q(.) = 0$ remains uniformly GES by invoking Lemma 4 in [10]. Hence, (38)-(39) can be viewed as a uniformly GES system perturbed by bounded perturbations, *i.e.*, $d(.)$ and $q(.)$. The proof is completed by invoking the converse Lyapunov theorem (such as Theorem 4.14 in [9], Lemma 1 in [10], and Theorem 3.3 in [11]), and the arguments known as total stability (such as Lemma 9.2, 9.3, Exercise 9.21, the last paragraph on page 349 in [9], and Theorem 24.1 in [12]). □

*Remark 4* From Proposition 3, we infer that the observer proposed in Proposition 1 works for any $\kappa > 0$ because the dynamics given in (11) and (27) are the same when $\rho$ equals zero. Additionally, if $\Psi$ equals zero, the observer proposed in Proposition 2 works for any $\kappa > 0$. Otherwise, since $\Psi(u)$ is a known function, $\kappa$ can be pre-set to satisfy (35). Regarding the stability analysis of Proposition 2, this process can even be done online. Specifically, we can define $\kappa(u)$ as a function of $\Psi(u)$ to satisfy (35) online and substitute it into (15) and (16) to form the dynamics of observer proposed in Proposition 2.

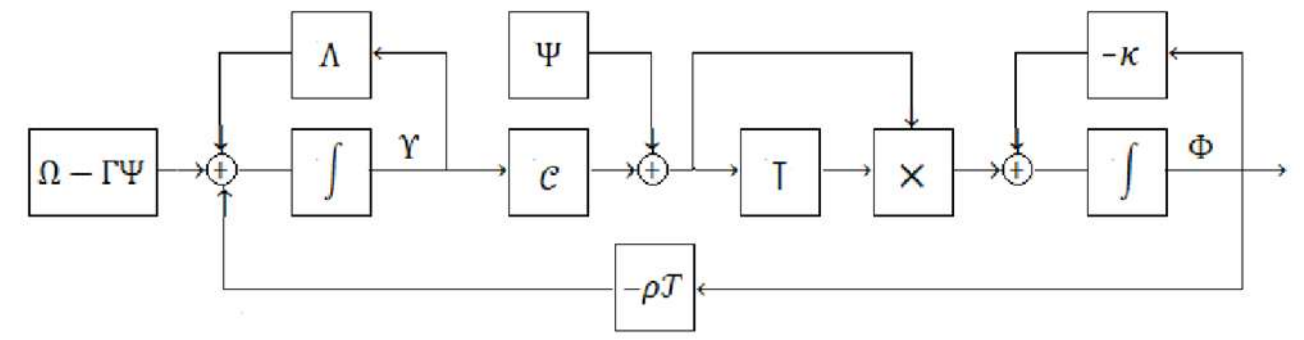

Fig. 1 Block diagram illustrating the connection between the dynamics of $\Upsilon$ and $\Phi$ in Proposition 1 ($\rho = 0$) and Proposition 2 ($\rho > 0$).

*Remark 5* As shown in Fig. 1, the dynamics of the dynamic regressor $\Upsilon$ and its extension $\Phi$ are connected in series in the DREM technique used in Proposition 1 and in [7]. In contrast, the approach proposed in Proposition 2 benefits from a feedback connection, which can be viewed as a nominal system with a damped nonlinear oscillator form, perturbed by some bounded perturbation terms (refer to the equations (36)-(42)). From the perspective of linear time-invariant systems literature, the feedback gain $\rho\mathcal{T}$ can generate complex conjugate poles for the nominal system in (38)-(39), causing the trajectories of $\Upsilon(t)$ and $\Phi(t)$ to oscillate in response to the perturbation terms. This transient behavior is instrumental in satisfying the parameter convergence conditions (18) and (19) within a finite time.

## V. Conclusion

The dynamic regressor extension and mixing procedure has been used to redesign a conventional adaptive observer algorithm for affine systems. A reduced-order observer was created without using the state transition matrix. The regressor's dynamics were restructured to benefit from feedback from its extension, transforming the regressor dynamics into a perturbed damped nonlinear oscillator form. This redesign allows for better parameter convergence and improves the excitation properties of the extension matrix. The next steps focus on the experimental validation of the method and its extension to the problem of model reference adaptive control.